\documentclass{sig-alternate}
\usepackage{graphics}
\usepackage{subfigure}
\usepackage{color}
\usepackage[pdftex]{hyperref}
\newcommand{\comment}[1]{}
\newcommand{\vkp}[1]{{{ #1}}}   

\definecolor{Orange}{rgb}{1,0.5,0}

\makeatletter
\let\@copyrightspace\relax
\makeatother

\begin{document}


\title{A Mobile Phone based Speech Therapist}
\numberofauthors{3}
\author{
  \alignauthor Vinod K. Pandey\\
    \affaddr{TCS Innovation Lab}\\
    \affaddr{Tata Consultancy Services Thane, India}\\
    \email{vinod.pande@tcs.com}
	\alignauthor Arun Pande\\
 \affaddr{TCS Innovation Lab}\\
    \affaddr{Tata Consultancy Services Thane, India}\\
    \email{arun.pande@tcs.com}
	\alignauthor Sunil K. Kopparapu\\
 \affaddr{TCS Innovation Lab}\\
    \affaddr{Tata Consultancy Services Thane, India}\\
    \email{sunilkumar.kopparapu@tcs.com}
}

\maketitle

\begin{abstract}
Patients with articulatory disorders often have difficulty in 
speaking.
These patients need 
several speech therapy sessions to enable them
speak normally. 
These therapy sessions are conducted by a specialized speech therapist.
The goal of speech therapy is to develop good speech 
habits as well as to teach how to articulate sounds the right way. 
Speech 
therapy is critical for continuous improvement to regain normal speech. 
Speech therapy sessions require a patient to travel to a 
hospital or a speech therapy center for extended periods of time 
regularly; this makes the process of speech therapy not only  time consuming 
but also very expensive. Additionally, there is a severe 
shortage of trained speech therapists around the globe in general and in 
developing countries in particular.
In this paper, we
propose a low cost mobile speech therapist, a system that enables speech 
therapy using a mobile phone 
which eliminates the need 
of the patient to frequently travel to a speech therapist in a far away 
hospital. The proposed system, which is being built, enables 
both synchronous and 
asynchronous interaction between the speech therapist and the patient 
anytime anywhere.

\end{abstract}




\section{Introduction}

Nearly $6\%$ of the population suffer from some kind of 
articulation and language disorders \cite{c1}. 
Articulation disorders can be due to cleft lip, cleft palate, 
cleft lip and palate, autism, cerebral palsy, hearing impairment to name 
a few. Several studies have been conducted to understand social, 
emotional and behavioral problems of these patients \cite{c2}, 
\cite{c3}, \cite{c4}. While 
surgery is not always required to correct 
articulatory defects, but almost in all cases the patient 
needs to undergo extensive speech therapy to rectify articulation 
disorder to regain normal speech.

Speech therapy is critical for continuous improvement in the patient 
speech. The goal of speech therapy is to develop good speech habits as 
well as to teach how to produce sounds correctly. Conventionally, speech 
therapy sessions are face to face and require a patient to travel to a 
hospital\footnote{at least in developing countries like India}
 or a speech therapy center for extended periods of time and 
frequently;
this makes the process of 
speech therapy not only  time 
consuming exercise in terms of travel for the young patient but also 
very expensive in terms of stay for extended duration near the speech 
therapy centers. Additionally, there is a severe shortage of trained 
speech therapists around the globe in general and in developing 
countries, like India, in particular. The severity of the problem 
increases for the rural patients affected by articulation disorders. Any 
form of speech therapy, which does not require frequent travel, but is 
monitored as effectively as a physical visit to a speech therapist is 
not only welcome but is an urgent requirement in all developing 
countries.

In this paper, a work in progress, 
we propose a system architecture  which could be used effectively 
by 
patients with articulation disorders to undergo speech therapy under the 
speech therapist guidance to regain normal speech. The approach takes 
advantage of the proliferation of the mobile phone and the advances in 
speech signal processing to facilitate patients undergo speech therapy 
sessions without (a) the actual physical presence of the speech therapist,
which essentially means the patient can avoid travel and (b)
the patient and the speech therapist need not be available on
this platform at the same time, meaning a speech therapist can assist a lot
more patients.
The proposed system also permits verbal and/or 
textual communication between remotely located patient and a speech 
therapist. The system uses the mobile phone to capture the speech of the 
patient and the analysis of the speech is done on a remote server. 
While this system is intended for speech therapy sessions, it can however
also 
be used for practicing speech lessons. The
proposed system addresses the issue of imparting speech therapy 
effectively, economically and remotely using available and to be 
developed state of the art technology. We present an approach 
to make speech therapy affordable and sustainable for use by masses. The rest
of the paper is organized as follows. 
In Section \ref{sec:lit} we browse
through related literature and describe our approach in Section
\ref{sec:proposed_approach}. In Section \ref{sec:analysis} we
discuss some experiments carried out on normal and speech with articulation
disorders. We discuss out interaction with speech therapist in
\ref{sec:discussions} and conclude in Section \ref{sec:conclusions}.

\section{Related Literature}
\label{sec:lit}

The issue of imparting speech therapy has been discussed in literature sparsely and
most of them are found in terms of patents. 
A miniaturized device for remote speech therapy to overcome stuttering problem
is reported in \cite{c5}. The described system acquires patients 
medical history and speech samples to perform automatic diagnosis of the speech 
disorder and connects with a therapist for teleconferencing. Based on the diagnostic information, patient can select a pre-programmed
 therapy session for speech training. The system also permits playing reference sound, storing patients speech samples and retrieving
 previously stored speech samples for play back and/or speech analysis.
This is
more of a web based system, which requires the speech therapist to be online
when the patient is using the system; further unlike what we propose, the
words that the patient is asked to speak is static and does not depend on the
progress of  the patient. 

A system for providing speech therapy and speech assessment by capturing tongue
movement of patient is described in \cite{c6}, however this is not intended for
conducting remote speech therapy.  
The system uses a special instrument, palatometer for acquiring labial, linguadental, lingapalatal contacts along with the voice samples. The 
system provides feedback on a PC, connected to the palatometer, by displaying contacts of lips, tongue etc for the spoken 
utterance. Simultaneously, lingual movements (contacts of lips, tongue etc) are
displayed on the screen for the reference speech. 
A scoring, based on the timing between the sensor and the lingual movement, is used to judge the closeness of the spoken 
utterance with the reference. 
In \cite{c7} a method for enhancing the fluency of persons who stutter while
speaking is addressed. They  use 
frames of eyeglasses
to  visually display 
the articulatory movements of a patients mouth and normal 
person's mouth to enable the patient observe the difference¿.
The patient can refer to the display at 
desired times to enhance the fluency of the speech. 
%
The system described in 
\cite{c8} presents to a user a symbol representative of a word, which the
patient has to to pronounce 
into a microphone, the therapist hears the pronounced word and enters
the phonetic 
representation of the user pronunciation. The system then 
automatically determining whether an error exists in 
the user pronunciation; and if an error exists, it categorizes the error. 
This system reported in \cite{c8} always requires the speech therapist to be
present when the patient is undergoing speech therapy.
Ratio between therapist to patients are very low and hence it may not always be feasible that a therapist is present 
when the patient is taking speech training sessions.

Movement of tongue is used for providing speech therapy and speech assessment
in \cite{c9}. The system uses a sensor plate, similar to \cite{c6},
for acquiring labial, linguadental, lingapalatal contacts 
along with the voice samples. The system determines a
set of parameters representing a contact pattern between the tongue and the
 palate during the patients utterance and compares 
them with a set of parameters from normal speaker in terms of
deviation and accuracy scores.
Additionally, the system also displays contact location between tongue/ palate
and sensor plate for the patient 
utterance and the normal speaker. 
Nasal airflow acquired simultaneously with the speech of the patient
is used in \cite{c10}. Variations in nasal 
airflow and speech signal with timestamps are used as a visual assessment of the patient condition by therapist. 
This approach is not automated and requires presence of the therapist for the analysis. 

Almost all the systems described in literature  specifically consider the fact
that the therapist and the patient be present at the same point of time even if
they are not in face to face contact. This can not address the issue of the lop
sided ratio in terms of therapy needing patients and the therapist. Our
approach is one of enabling a speech therapy session happen without requiring
the patient and the therapist to be present at the same time \cite{pat:rst}.

\section{Proposed Solution Approach}
\label{sec:proposed_approach}

It is well known that for producing different speech sounds, 
articulators such as jaw, tongue, teeth, lips etc move from one position 
to another. The configuration of these articulators effectively change 
the shape of the vocal tract which usually ends at the lips for some 
sounds and at the nose for some sounds. This position change in 
articulators is exhibited in some form in the acoustic signal produced
by the human speech.
Typically a speech 
signal is represented by several parameters such as energy, pitch, 
formats, jitter, shimmer, spectral tilt, spectral balance, 
spectral moments, spectral decrease, spectral slope and its roll-off.
 
 Different types of articulation disorder may need 
different dictionary words, phrases and sentences that need to be practiced by
the patient. 
For example, therapy in case of
cleft patients may involve speaking more 
words with nasal sounds. 
A database of words, phrases and sentences which have sounds 
embedded in them which are typical of being wrongly produced by patients 
with certain articulation disorders is maintained. 
These words, 
could also be phrases or sentences, are (a) chosen in 
consultation with the speech therapist along with (b) details of the order in 
which these words need to be presented to the patient and additionally 
(c) the 
manner in which the spoken words are to be analyzed to identify the 
improvement in articulation post a speech therapy session.  
Typically references of normal speech is 
stored in terms of the speech parameters such as energy, pitch, formats,
 jitter, shimmer, spectral tilt, spectral balance, 
spectral moments, spectral decrease, spectral slope and its roll-off etc 
in a database. These speech features can be used to compute 
the closeness of the patient speech and the reference speech. This 
closeness metric can also be also used to (a) judge the progress of the patient
and (b)
as a  guide during speech therapy session.

\begin{figure}
\centerline{\includegraphics[width=0.60\textwidth]{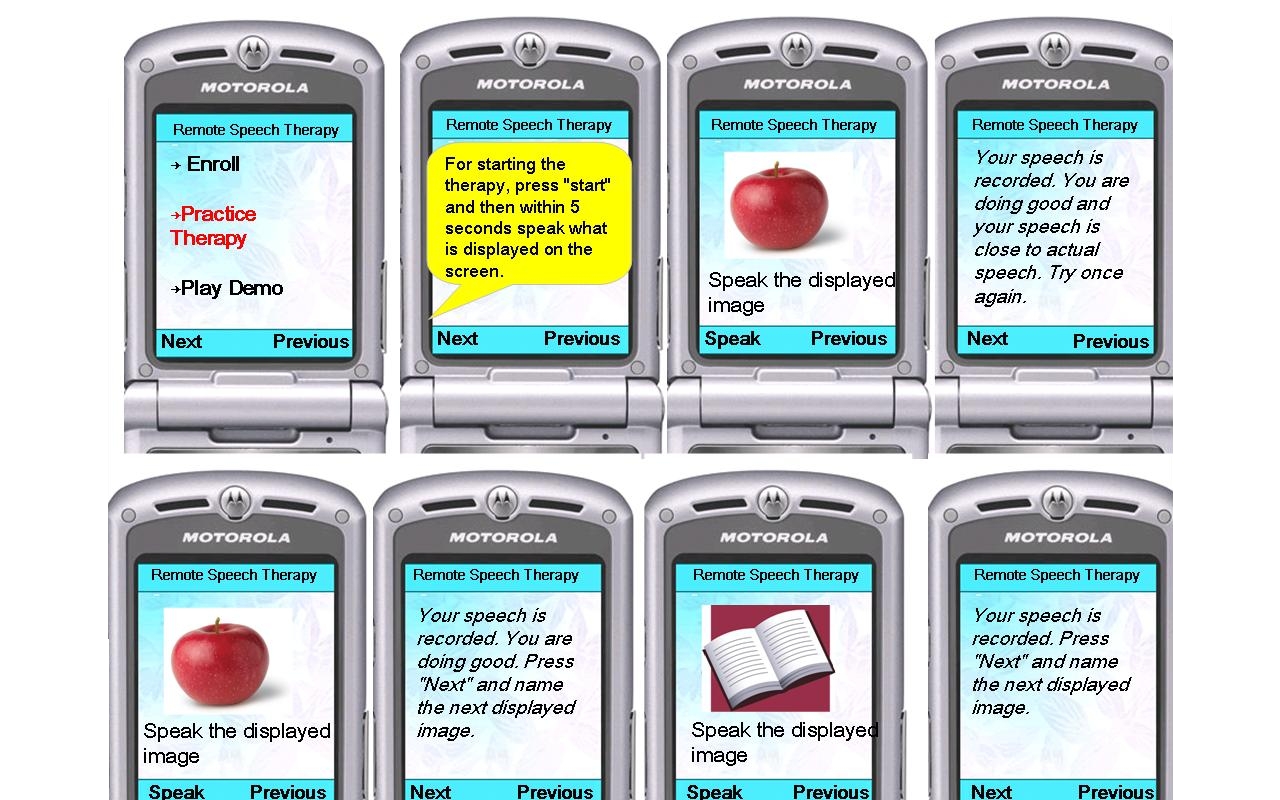} }
\caption{Instances of the application running on patient mobile phone.}
\label{fig:Fig2}
\end{figure}

 
At a very high level 
architecture of the proposed system is one of a web service. The 
system connects the patient and the speech therapist without both of them
being
available on the system at the same time. 
A typical scenario would require the mobile phone, with the
proposed remote speech therapy application, be with 
the patient who wishes to undergo speech therapy. 
Fig. \ref{fig:Fig2} shows 
instances of the application running on the patients mobile phone. 
As is normal in any hospital, 
there is a registration process where in several details of 
the patient are captured, like name, age, gender, medical history, 
nature of articulatory disorder, name of the speech therapist etc. 
Registration details serve two purposes, (a) it determines the speech therapist
who can view the progress of the patient and (b) determine the nature of
therapy that is required for the patient. 
The application on the mobile
phone is controlled by the server 
to utter a certain word or a phrase. 
The set of words and the sequence in which they need to be presented to the
patient is controlled by the server and is 
based on the articulatory disorder of the
patient. 
Additionally, the set of words and the sequence of the
words that need to be uttered by the patient can be changed, if desired, 
by the speech therapist on the server offline. 
This word the patient has to speak is elicited by presenting 
an image, displayed on the mobile screen or as a spoken speech sample. 
The patient is expected to speak the word when prompted by the system. 
The patient 
spoken speech is then processed on the server and then 
compared with the reference normal spoken speech. The method of 
comparison is based on what a therapist would normally look for in a face to
face session. For example, if a certain vowel in a word  is what is to be
observed then the comparison would be based on the 
identification of the vowel in
the spoken speech followed by vowel comparison with reference vowel.
The comparison,  
enables identification of the disorder and  in case of post speech
therapy the degree of improvement in speech. 
The deviation, if any, from the normal
spoken speech is 
presented as a feedback to the patient. 
The feedback could be in
the form of 
either (a) 
graph and presented to the patient in a manner that might assist the patient to
improve the articulation and thereby rectify the disorder or 
(b) as a reference sound to enable the patient 
to {\em learn} the sound.  This process of the patient uttering a sound and 
the system identifying the existence of the disorder and displaying a 
feedback to assist the patient is repeated with out the actual presence
of the speech therapist on the other side of the system at the same time. 

\begin{figure}
\subfigure[Patient Details.]
{ \includegraphics[width=0.50\textwidth]{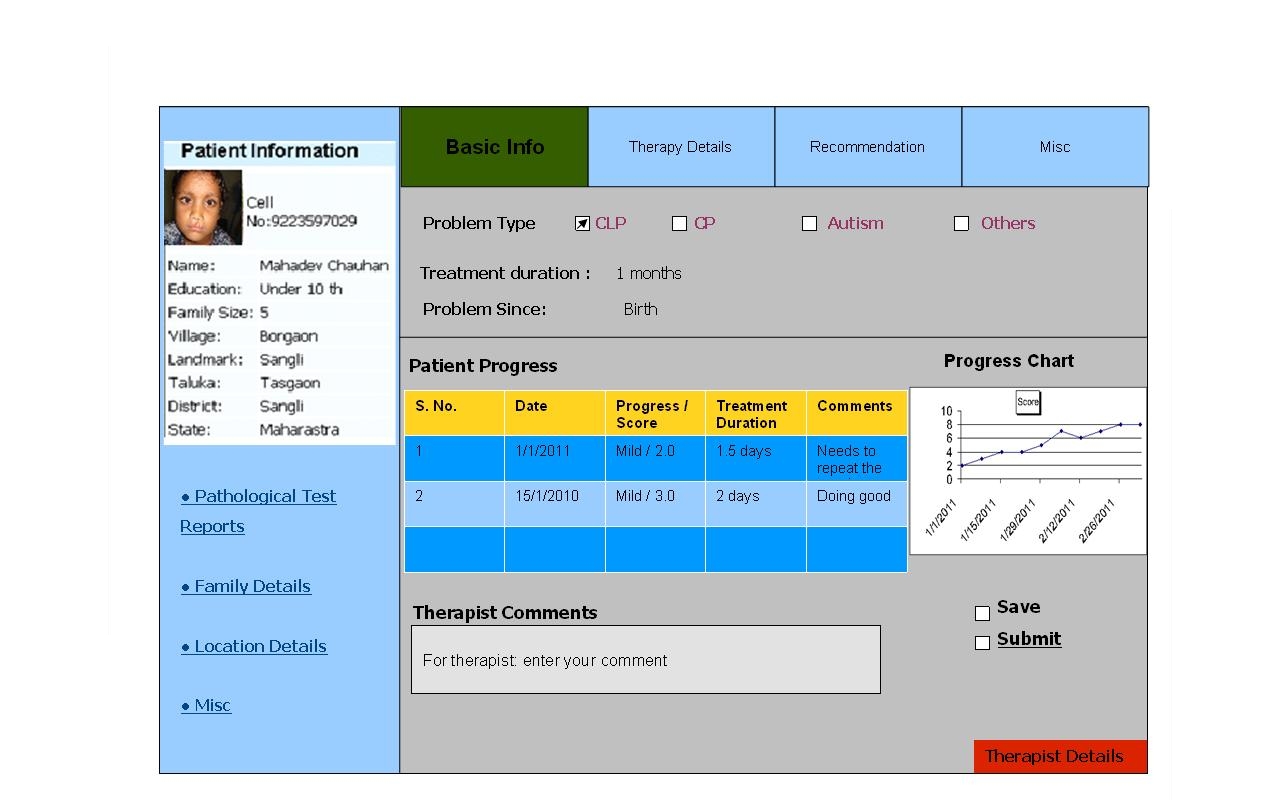}
\label{fig:Fig3}
}
\subfigure[All Therapy Sessions.] 
{ \includegraphics[width=0.50\textwidth]{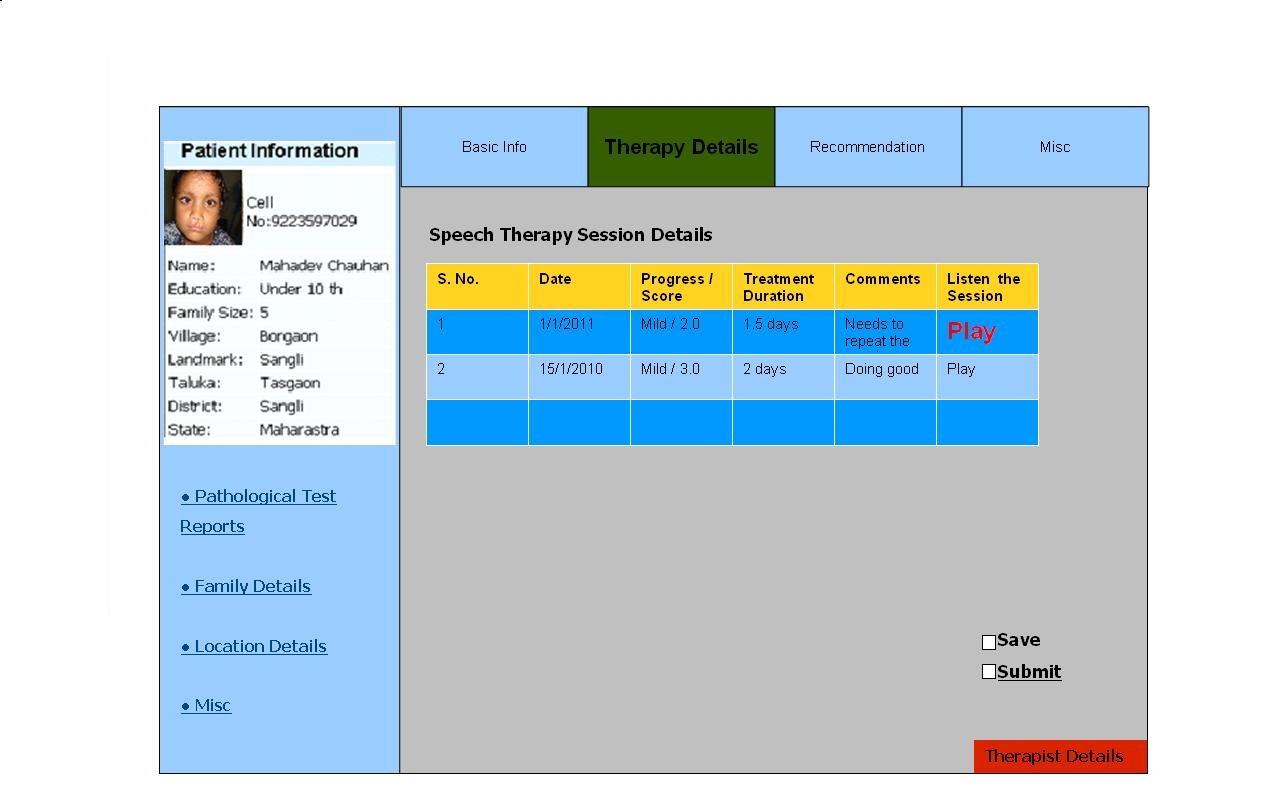}
\label{fig:Fig4}
}
\subfigure[Particular Therapy Session.]
{ \includegraphics[width=0.50\textwidth]{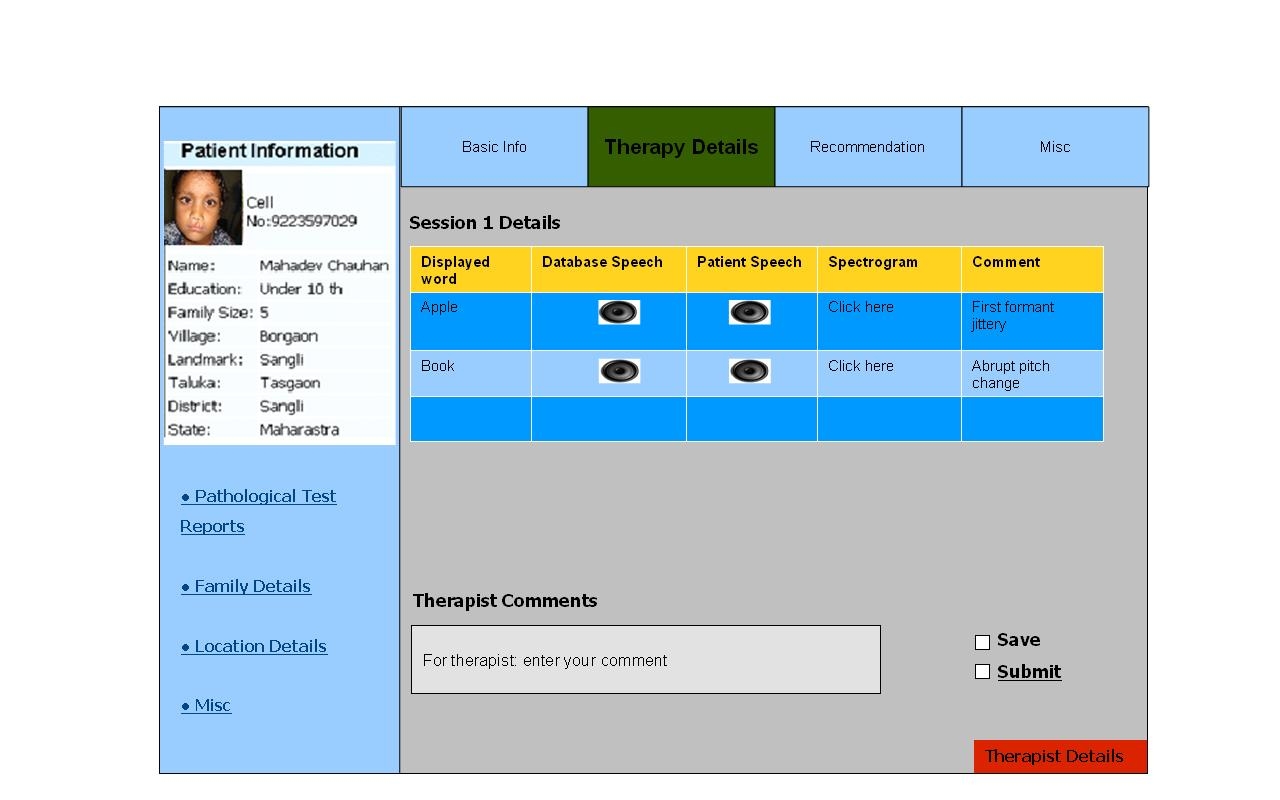}
\label{fig:Fig5}
}
\caption{Speech Therapist Console.}
\label{fig:console}
\end{figure}

The speech therapy sessions taken by the patient can be examined 
at the therapist convenience periodically.
Patients speech files and other performance details of the patient are
presented as 
a dash board view (an expert console) for the therapist to browse the activity
of the patient.
therapist on 
using a personal computer connected to the Internet.
Fig. \ref{fig:console} 
shows different instances of the dash board as viewed by the therapist.
Fig. \ref{fig:Fig3} is the speech therapist view which 
displays details such as personal information of 
patient, nature and date of surgery, 
speech therapy sessions taken by the patient and the current state of the 
patient responding to the speech therapy, Fig. \ref{fig:Fig4} shows details of
all the speech therapy sessions taken by the patient while Fig. \ref{fig:Fig5}
shows a view where in the speech therapist can play the actual speech uttered
by the patient and review the performance.
Additionally the speech therapist can get to view the amount of time 
spent by the patient in training or using the speech therapy system. 
This dash board would evolve with discussions with speech therapist and would 
strive to give a good 
picture of the progress made by the patient as desired by the speech therapist. Speech therapist can also 
use this dash board to communicate offline with the patient if desired; typically as 
a voice instruction, which the patient will hear when he uses his mobile phone
to take his/her next therapy session.

The proposed system facilitates both real time online communication 
and offline communication between the patient and the speech therapist. 
In an online mode, a human speech therapist is available during the speech 
therapy session, guiding the patient in real time. This is more like a remote
speech therapy, except that the patient and the therapist are not at the same
location. However, in an offline mode, 
the word uttered by the patient and the reference word stored on the server
are used for comparison and giving automatic feedback. In this scenario the
patient and the therapist need not be present on the system at the same time.
Details of the online and offline speech therapy sessions along with the spoken speech samples are stored on the server and are available 
for a speech therapist to review 
and take necessary corrective measures to speed up or slow down the therapy sessions.

\section{Preliminary Analysis}
\label{sec:analysis}

\begin{figure}
\centerline{
\includegraphics[width=0.50\textwidth]{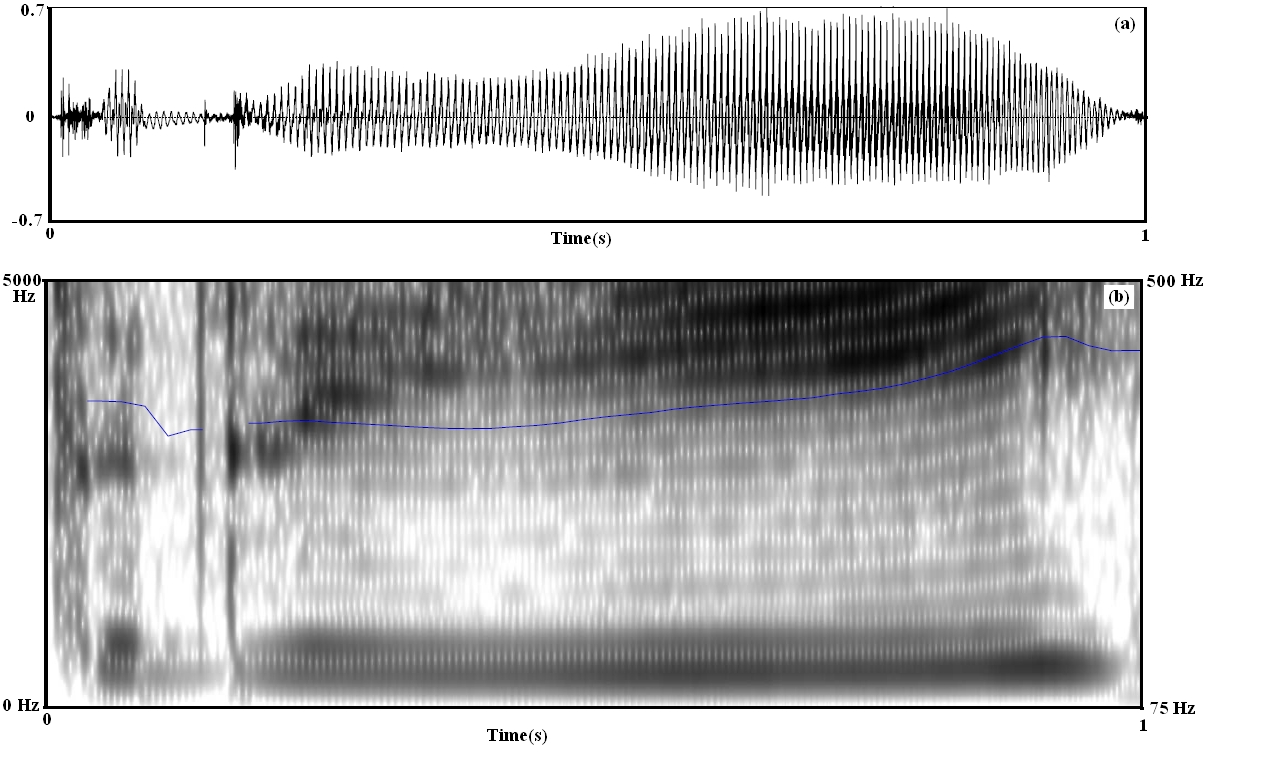}
}
\caption{(a) Speech waveform, (b) spectrogram and corresponding pitch contour,
for English numeral {\em /three/}, spoken by a normal female subject (age: 6 years). Time axis: normalized to 1.}
\label{fig:Fig6}
\end{figure}

\begin{figure}
\centerline{
\includegraphics[width=0.50\textwidth]{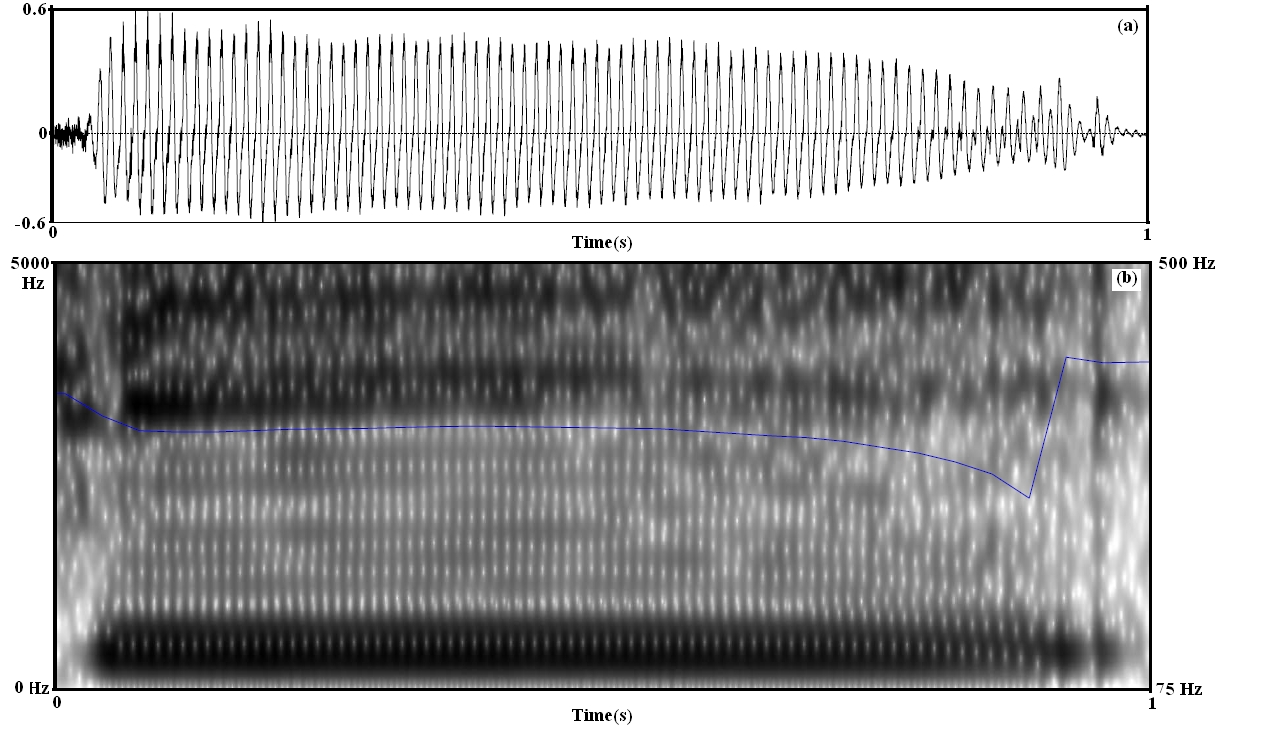}
}
\caption{
(a) Speech waveform, 
(b) spectrogram and corresponding pitch contour,
for English numeral {\em /three/}, 
spoken by a CP female subject (age: $5.8$ years). 
Time axis: normalized to $1$.}
\label{fig:Fig7}
\end{figure}

We conducted some preliminary experiments to analyze if there were identifiable
differences between normal and a person with articulation disorder. 
Speech signals from normal speaking and cleft palate children were 
analyzed to observe visible differences in their characteristics. 
A total of four 
speech samples ($2$ normal and $2$ cleft palate) were analyzed; the spoken 
speech was that of English numerals from {\em one} to {\em ten}. 
A visual difference 
between normal and cleft palate speech especially for some speech 
features, namely, pitch contour, formant, and signal intensity was 
observed. Fig \ref{fig:Fig6} shows speech waveform, spectrogram, and 
pitch contour for the normal female subject for spoken utterance {\em 
/three/}. Fig \ref{fig:Fig7} shows the speech waveform, spectrogram, and 
pitch contour for a female subject with  cleft palate (CP) 
for the same spoken utterance {\em /three/}. 
Visual observation 
of some features shows that 
\begin{itemize}
\item the 
pitch contour for the CP female subject has abrupt changes, specially 
for spoken word: {\em /three/} and {\em /four/}. 
\item For the male group (age: 6 years), the 
pitch contour for the normal male subject was more continuous than the pitch 
contour for the CP affected male subject, which had a lot of discontinuities. 
\item Duration of the 
spoken numeral was smaller for the CP female subject compared to the normal
speech. 
\item However, 
there was no noticeable difference in duration of spoken words for the 
normal male and CP male subject. 
\item Mean value of the first 
formants for the CP male and female subjects were lower as 
compared to the mean of the first formant for the normal male and female 
subjects. 
\end{itemize}
Some of these features can be used for discriminating normal speech and 
 speech from patients with articulatory disorder. 

\section{Discussion}
\label{sec:discussions}

While some observation can be derived from the small set of analyzed 
data, this 
in no way suggests that (a) these observations are comprehensive, 
(b) these speech feature set are complete and sufficient. We are in 
process of recording large speech data from different patients with 
different articulation disorder.
It is proposed to
identify a small set of articulatory disorders and set up a complete 
speech therapy session in consultation with a speech therapist. We
plan to run a pilot by distributing the mobile 
based application to some patients of the 
institute for the hearing handicapped to carry out 
speech therapy remotely. 
The patient will be directed to speak certain words,
and the presentation of the next word or phrase will be based on the closeness
of the patient spoken word to the normal spoken word. If the patient spoken
word is articulated wrongly, the system would prompt the patient to repeat the
word by giving a feedback on where s/he 
misarticulated. All
therapy session will be logged and the speech files stored on the
server.
 The dash board will enable the 
speech therapists to retrieve and listen to the patient speech 
samples, and also see the analysis results at a later time convenient to the
speech therapist. The dash board will also provide
facility to the speech therapist to modify the therapy program (sequence of
words to be presented to the patient) depending
on the speech therapist analysis of the progress of the patent; this provides 
the
speech therapist an option to override the system analyzed option.
The actual success of the proposed remote speech
therapy system will be based on the
feedback provided by the patients and also the speech 
therapists which will give an idea if this method of speech therapy can reduce
the number of visits of the patient to the speech therapist thereby making the
remote speech therapy (a) cheap and making speech therapy accessible to more
patients. We also plan to monitor the reduced number of trips to the therapist
and the reduced time and travel cost.

\subsection{Discussion with Speech Therapist}

Initially, we approached 
a hearing impaired institute in Mumbai and spoke to several 
speech therapist in the institute
and suggested the concept of developing a mobile phone based remote speech 
therapy aid. During initial interactions, 
the therapists had hesitation in viability of the tool for
speech therapy. 
They were convinced with the proposal after a short demonstration  
of the proposed system. This helped us in being able to get audience 
with 
the patient when the speech therapy was in progress. Several observations that
will be of use to build the system came up. Like,
a therapist initially 
checks if the patient is familiar with sustained sounds 
(like {\em 't'}, {\em 'd'}, {\em 'k'} and {\em 'p'} etc), 
commonly used words, counting numbers, phrases 
(mostly stories, rhyme etc) before prescribing a therapy.
A therapist often concentrates on speech parameters like pitch, 
loudness, and frequency (spectrogram) for differentiating patient speech
and normal speech. 

The therapists 
were of the 
opinion that the proposed system can definitely reduce the number of 
visit of the patient to the therapy centers.
Personalization and design flexibility of the dictionary for different
languages was another aspect that came out. 
The therapists suggested that it might be meaningful to have a facility on the
proposed platform that can enable
communication with the patients, as a SMS or MMS for giving instructions and 
prescriptions, replying to patients query.
They felt that the  remote speech therapy system could possibly enhance the
pace of the therapy.

\section{Conclusions}
\label{sec:conclusions}

Patients with articulation disorders needs guidance from a speech 
therapist to regain their normal speech. The ratio of the speech therapist and
the patients that need speech therapy is poor. Several visits to actual speech
in progress therapy session suggests that it is feasible to provide a platform
that can bridge the gap between the number of patient that need speech therapy 
and the speech therapist by reducing the actual number of
physical visits of the patient to the speech therapist. This would enable the
speech therapist to {\em see} more patients.
The central idea of the 
proposed system is to enable speech therapy with out requiring a speech 
therapist being physically available during an actual in progress
 therapy session. This 
would in many cases result in avoidance of the need of the patient to 
travel to a speech therapist at a distant location / hospital. In brief, 
this mobile based system will enable putting the patient and the speech 
therapist in a virtual room though they are physically separated 
geographically.  With appropriate user interface and tested usability of 
the software, mobile phone can be a robust device in the hands of the 
patients to undergo speech therapy sessions and interact with the 
therapist remotely. Patients can take speech therapy sessions at their 
convenience. This will be useful specially for young kids. This device 
can also be used to display multi media information in the form of audio 
clips, video clips for the purpose of training.

\section*{Acknowledgment}

Several members of the TCS Innovation Labs - Mumbai have been instrumental in
contributing to the conceptualization and building the system. Special thanks
to 
faculty members and speech therapist at Ali Yavar Jung National 
Institute for the Hearing Handicapped, Mumbai, India for simulating 
discussions which enhanced our understanding of 
how speech therapy actually happens on ground.

\end{document}